\documentclass[prl,twocolumn,twoside,showpacs,superscriptaddress]{revtex4}
\usepackage{graphics,amssymb,amsmath,epsf}
\epsfclipon

\begin{document}
\title{Entrainment transition in populations of random frequency oscillators}

\author{Hyunsuk Hong}
\affiliation{Department of Physics and RINPAC, Chonbuk National University,
Jeonju 561-756, Korea}

\author{Hugues Chat\'e}
\affiliation{Service de Physique de l'Etat Condens\'e, CEA-Saclay, 
91191 Gif-sur-Yvette, France}

\author{Hyunggyu Park}
\affiliation{School of Physics, Korea Institute for Advanced Study, 
Seoul 130-722,
Korea}

\author{Lei-Han Tang}
\affiliation{Department of Physics, Hong Kong Baptist University, 
Kowloon Tong, Hong Kong SAR, China}

\date{\today}

\begin{abstract}
The entrainment transition of coupled random frequency oscillators is 
revisited. The Kuramoto model (global coupling) is shown to exhibit 
unusual sample-dependent finite size effects leading to a correlation
size exponent $\bar\nu=5/2$.
Simulations of locally-coupled oscillators in $d$ dimensions
reveal two types of frequency entrainment: mean-field behavior at $d>4$,
and aggregation of compact synchronized domains in three and four dimensions.
In the latter case, scaling arguments yield a correlation length exponent
$\nu=2/(d-2)$, in good agreement with numerical results.
\end{abstract}
\pacs{05.45.Xt, 05.45.-a, 89.75.-k}
%05.45.Xt: Synchronization, coupled oscillators
%05.45.-a Nonlinear dynamics and nonlinear dynamical systems
%89.75.-k: Complex systems
\maketitle

Collective oscillations are abundant in physical, chemical, and biological
systems far from 
equilibrium~\cite{ref:Winfree,ref:Kuramoto,ref:Pikovsky,ref:Barbara,ref:Kiss}.
Such synchronized behavior has been widely explored via various systems of
coupled oscillators~\cite{ref:Kuramoto1,ref:additional,ref:Strogatz,ref:Acebron}.
Although several theoretical methods are available to treat systems of identical
oscillators with or without noise~\cite{ref:Cross,ref:GMSB,ref:DeMonte},
the description of synchronization in coupled oscillators with a broad
distribution of intrinsic frequencies is still a largely unsolved
problem~\cite{ref:Strogatz,ref:Acebron}.  More recently, the study of dynamics
on complex networks, including the human brain, brought renewed interest
into the subject~\cite{ref:complex_net}.

The onset of system-wide synchronization at some coupling strength is often
compared to critical phenomena in equilibrium systems, where scaling concepts
provide an adequate theoretical basis for quantitative analysis.
The scaling approach has been successfully applied to
the Kuramoto model~\cite{ref:Kuramoto1},
but its extension to finite-dimensional systems has stumbled over a number of
difficulties~\cite{ref:Sakaguchi,ref:Daido88,ref:Strogatz88,ref:Hong}.
Even in the globally-coupled case, 
the critical size needed to maintain a coherent
group of synchronized oscillators in a large but finite population,
as well as their ensuing dynamics, have not been fully
understood~\cite{ref:Strogatz,ref:Acebron,ref:Hong}.
In $d$-dimensional systems with local coupling, $d=2$ is generally
accepted as the lower critical dimension for macroscopic entrainment,
but the upper critical dimension is uncertain due to
peculiar size effects seen in numerical investigations~\cite{ref:Hong}.

In this Letter, we re-examine the above issues critically and establish
two types of behavior at the onset of macroscopic synchronization in
coupled random frequency oscillators. In the globally coupled case
and in finite dimensions $d>4$, we show that frequency entrainment is
accompanied by spontaneous symmetry breaking in the phase, and as such
shares many features with the usual ordering transition in equilibrium
systems with $O(2)$ symmetry. The correlation size exponent $\bar{\nu}$
governing finite-size scaling at criticality, however, takes on the value 
$\frac{5}{2}$ in the present case rather than 2 in usual 
mean-field theories~\cite{ref:Botet}.
For $d\leq 4$, the above type of symmetry-breaking
is ruled out by diverging phase fluctuations. Instead, oscillators
are first entrained with their neighbors to form compact, synchronized domains
which grow upon further increase in coupling. 
System-wide entrainment is achieved
only when the largest domain reaches system size. This gives rise to
size- and sample-dependent thresholds whose behavior is quantified.

We begin with the globally-coupled Kuramoto model defined by the following
set of dynamical equations:
\begin{equation}
{d\phi_j\over dt}=\omega_j-K\Delta\sin(\phi_j-\theta).
\label{mf-phi}
\end{equation}
Here $\phi_j$ is the phase of the $j$th oscillator ($j=1,\ldots,N$),
and $\omega_j$ its natural frequency, drawn from a normalized distribution 
$g(\omega)$. To be definite, here we take $g(\omega)$ to be Gaussian
with zero mean and unit variance. The parameter $K$ sets the strength
of coupling to the global quantities $\Delta$ and $\theta$ defined via
\begin{equation}
\Delta e^{i\theta}=N^{-1}\sum_{j=1}^N e^{i\phi_j} \;.
\label{order-parameter}
\end{equation}

In the classic work by Kuramoto, the existence of an entrained state 
is established
by considering solutions to (\ref{mf-phi}) at a constant $\Delta$. After an
initial transient, oscillators with $|\omega_j|<K\Delta$ reach a fixed angle
$\phi^{(0)}_j$ while those with $|\omega_j|>K\Delta$ are in a running pendulum
state. In the limit $N\rightarrow\infty$, the self-consistent equation for
$\Delta$ reads
\begin{equation}
\Delta=\int_{-K\Delta}^{K\Delta}d\omega g(\omega)\sqrt{1-(\frac{\omega}{K\Delta})^2}
\equiv\Psi(\Delta) \;,\label{Delta-mf}
\end{equation}
which has a nontrivial solution when $K>K_{\rm c}=\frac{2}{\pi g(0)}$.
For a population of finite size, $\Psi(\Delta)$ is replaced by
\begin{equation}
\tilde\Psi(\Delta)\equiv {1\over N}\sum_{j,|\omega_j|<K\Delta}
\sqrt{1-\Bigl({\omega_j\over K\Delta}\Bigr)^2},
\label{Delta_tilde}
\end{equation}
which contains a sample-dependent correction
$\delta\tilde\Psi\equiv\tilde\Psi-\langle\tilde\Psi\rangle
=\tilde\Psi-\Psi \propto\sqrt{N_{\rm s}}/N$. 
Here $N_{\rm s}$ is the number of oscillators in the frequency interval
$(-K\Delta,K\Delta)$, and $\langle\cdot\rangle$ denotes sample average. Close
to the transition, the self-consistency equation
$\Delta=\tilde\Psi(\Delta)$ can be written as
\begin{equation}
\Delta = aK\Delta - c(K\Delta)^3+\delta\tilde\Psi,
\label{delta_N}
\end{equation}
where $a=K_{\rm c}^{-1}$ and $c=-\pi g''(0)/16$. The variance of
$\delta\tilde\Psi$ is given by
$\langle(\delta\tilde\Psi)^2\rangle=\frac{4}{3}g(0)K\Delta/N+O(\Delta^2/N)$. 
Hence the solution to Eq.~(\ref{delta_N}) takes the scaling form
\begin{equation}
\Delta(K,N)=N^{-1/5}f(kN^{2/5}),
\label{delta-scaling}
\end{equation}
where $k\equiv (K-K_{\rm c})/K_{\rm c}$.
% is the distance to the transition.
The scaling function $f(x)$ is sample-dependent
and satisfies the equation,
\begin{equation}
xf-cK_{\rm c}^3f^3+(8/3\pi)^{1/2}\epsilon f^{1/2}=0.
\label{f-eqn}
\end{equation}
Here 
$\epsilon\equiv\delta\tilde\Psi/\langle(\delta\tilde\Psi)^2\rangle^{1/2}$
is a Gaussian random variable with zero mean and unit variance.

\begin{figure}
\epsfxsize=\linewidth
\epsfbox{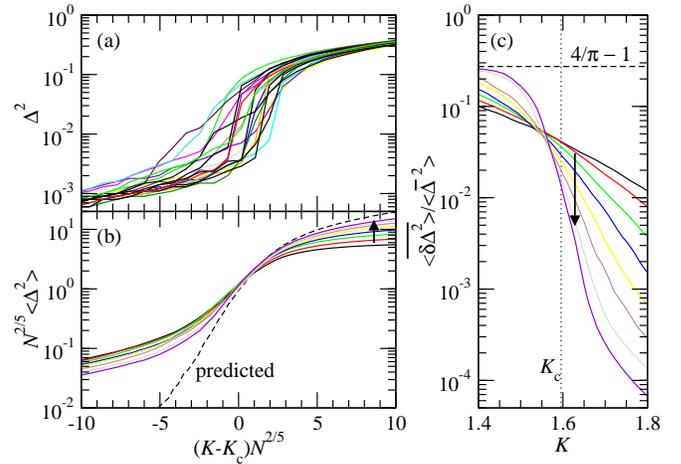}
\caption{ (Color online)
$\Delta^2$ against $K$ for globally-coupled
oscillators. (a) Time-averaged value of $\Delta^2$ for 20 different 
samples ($N=12800$). (b) Scaling plot
for $N=100, 200,\ldots, 12800$, in increasing order
as indicted by the arrow (averaged over 100 samples for each $N$).
Dashed line: see text. (c) Relative strength of temporal versus
sample-to-sample fluctuations of $\Delta$ for a set of system sizes as in (b).
}
\label{fig:mf-delta_scaling}
\end{figure}

Equation (\ref{delta-scaling}) resembles the finite-size scaling form in
various mean-field models of equilibrium 
phase transitions~\cite{ref:Botet}, but
the value of the exponent $\bar{\nu}=\frac{5}{2}$ is quite unusual. 
To check that $\Delta$ in the transition region is dominated by
$\delta\tilde\Psi$ arising from ``density fluctuations'' of oscillators
along the frequency axis, we have performed
extensive simulations of Eq.~(\ref{mf-phi}). 
Figure~\ref{fig:mf-delta_scaling}(a)
illustrates sample-to-sample fluctuations in the time-averaged value of
$\Delta^2$. 
The onset of entrainment spreads over a distance $\delta K\sim N^{-2/5}$ 
around the nominal $K_{\rm c}$,
as seen from the scaling plot of $\langle\Delta^2\rangle$ against $K$ 
at various sizes $N$ in Fig.~\ref{fig:mf-delta_scaling}(b).
The dashed line there is obtained by averaging solutions to Eq.
(\ref{f-eqn}) at many different values of $\epsilon$ drawn from a Gaussian
distribution. The agreement between the predicted scaling and simulation data
is satisfactory on the entrained side and sufficiently close to the transition,
on which Eq. (\ref{delta_N}) is based.
Figure~\ref{fig:mf-delta_scaling}(c) 
shows the relative strength of {\it temporal} fluctuations
$\delta\Delta(t)\equiv \Delta(t) -\overline{\Delta}$
of the order parameter in the transition region. 
(Here the overline bar denotes time average.)
On the detrained side, the ratio 
$\langle\overline{\delta\Delta^2}\rangle/\langle\overline{\Delta}^2\rangle$
at large $N$ approaches the value $\frac{4}{\pi}-1$ for Gaussian fluctuations.
On the entrained side, this quantity decreases
as $N^{-1}$ as expected from independent fluctuations.
At $K=K_c$, the data clearly decrease with $N$, 
lending further support to our analysis.

We now turn to locally-coupled oscillators on $d$-dimensional 
hypercubic lattices,
for which the dynamical equations take the form
\begin{equation}
\frac{d\phi_i}{dt} = \omega_i - K\sum_{j\in \Lambda_i}\sin(\phi_i - \phi_j).
\label{eq:model}
\end{equation}
where $\Lambda_i$ is the set of all nearest neighbors of site $i$.

At sufficiently large $K$, Eq. (\ref{eq:model}) can be treated using a linear
approximation~\cite{ref:Sakaguchi,ref:Hong}.  
In this limit, the system enters a
``fully entrained'' state where the random term $\omega_i$ is balanced by local
gradients of a static phase field $\phi^{(0)}_i$. Salient features of this
state are summarized as follows: i) For $d>4$, $\phi^{(0)}_i$ has bounded
variations even when the linear system size $L\rightarrow\infty$. Consequently,
the entrained state has a global phase that breaks the $O(2)$
symmetry. ii) For $2<d\leq 4$, variations in $\phi^{(0)}_i$ grow as
$L^{(4-d)/2}$. Hence an entrained state {\it cannot} 
be assigned a definite phase.
iii) For $d\leq 2$, local phase gradients have an infra-red divergence
$L^{(2-d)/2}$. In this case, the system is detrained beyond a coherence length
$\xi\sim K^{2/(2-d)}$.

As $K$ decreases, the local phase gradients need to increase to counter
the $\omega_i$'s but there is a limit to how far this can go
in the Kuramoto model. Upon proliferation of phase slips and runaway
oscillators, two scenarios can be contemplated for the destruction of
global entrainment:
i) Oscillators break away individually from the entrained group 
until the latter is reduced to an infinitesimal fraction of the system
as in the globally-coupled model.
ii) A more collective form of phase slips takes place along 
``domain boundaries'' that break the system into locally synchronized 
clusters, starting from the largest scale.
For ii) to preempt i), there must be pre-existing large phase differences
(i.e., ``strain'') across the system, which is the case in the entrained 
state at and below $d=4$ but not above. This suggests that the nature
of the detrainment transition is quite different above and below $d=4$.

To characterize system-wide coherent phase motion in the entrained state,
we introduce the Edwards-Anderson order parameter
\begin{equation}
\Delta_{\rm EA}=\lim_{t-t_0\rightarrow\infty}{1\over N}
\Bigl|\sum_je^{i[\phi_j(t)-\phi_j(t_0)]}\Bigr|.
\label{EA}
\end{equation}
Compared to $\Delta$ defined by Eq. (\ref{order-parameter}),
$\Delta_{\rm EA}$ is incensitive to the static phase deformation 
discussed above, and hence can also be used for $d\leq 4$.

\begin{figure}
\epsfxsize=8 truecm
\epsfbox{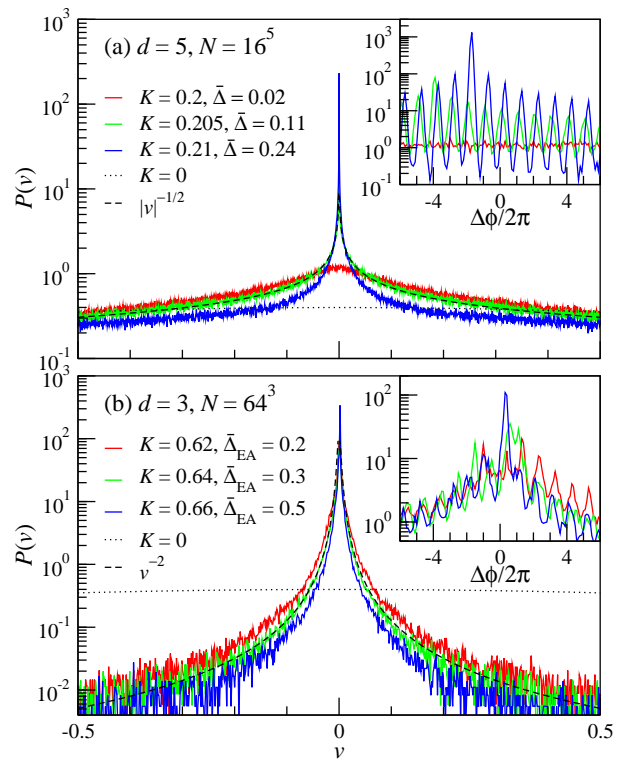}
\caption{(Color online) 
Distribution of phase velocities at various values of the coupling
strength $K$ in (a) five, (b) three dimensions. Insets
show the peak region in detail. 
The time-averaged value of the order parameter are listed.}
\label{fig:pdf_omega}
\end{figure}

We have carried out extensive numerical simulations of (\ref{eq:model})
to investigate the two types of transition behavior and 
associated critical properties and finite-size effects.
Periodic boundary conditions are used.
Let us first examine the spatio-temporal behavior of
the phase advance $\Delta\phi_i=\phi_i(t+T)-\phi_i(t)$ of oscillators
over a sufficiently long interval $T$ in a
given oscillator population near the entrainment transition.
Figure~\ref{fig:pdf_omega}(a) presents the distribution $P(v)$ of 
the mean phase velocity $v_i=\Delta\phi_i/T$ over an interval
$T=10 000$ for a $N=16^5$ system in $d=5$ dimensions. The inset 
shows a magnified plot of the peak region where values of
$\Delta\phi$ within each $2\pi$ interval are resolved. 
Evidently, entrainment here is accompanied by the selection of a global phase,
i.e., the transition is of the symmetry-breaking type. Comparing the
two distributions at $K=0.2$ and $K=0.205$, we see that only a small
fraction of oscillators participate in the onset of entrainment, while
the rest are not much affected at this stage. This
is very similar to the behavior of the globally-coupled case.
At $K=0.205$, the wings can be fitted
well to a weak, integrable power-law, $P(v)\sim |v|^{-1/2}$.

Figure~\ref{fig:pdf_omega}(b) shows $P(v)$
for a $d=3$ system of $N=64^3$ oscillators at three different
values of $K$ around the entrainment transition. Here $T=5 000$.
A strong narrowing of $P(v)$ is seen in the entire critical region, 
indicating the formation of large synchronized domains well below 
global entrainment. The wings fall off roughly as $v^{-2}$. Consequently, most
oscillators are moving at phase velocities close to that of the peak. 
However, as seen in the inset, the actual phase 
advance $\Delta\phi_i$ of these oscillators is much less entrained as 
compared to the $d=5$ case.

\begin{figure}
\epsfxsize=\linewidth
\epsfbox{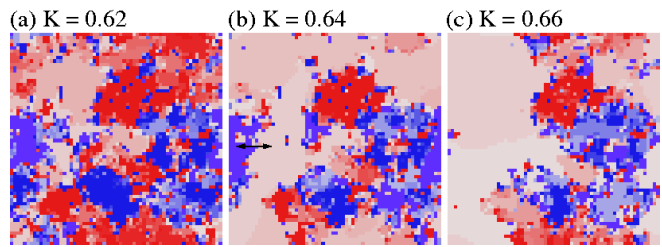}
\caption{(Color online) 
Spatial structure of $\phi_i(t+T)-\phi_i(t)$ in one layer from the
sample depicted in Fig. \ref{fig:pdf_omega}(b). Here $T=5000$.
Color scale from blue: $-12\pi$ to red: $12\pi$.
}
\label{fig:layer_3D}
\end{figure}

Further indication of entrainment through growth and aggregation of locally
synchronized domains is found in the spatial structures of the
$\Delta\phi_i$'s, as depicted in
Fig.~\ref{fig:layer_3D} for one layer of the $d=3$ system.
At $K=0.66$, a large, spanning domain of entrained oscillators coexist with 
smaller clusters of oscillators at varying phase velocities. 
When $K$ is decreased to 0.64, part of this largest domain undergoes
a phase slip over the time period, as indicated by the arrow
in Fig.~\ref{fig:layer_3D}(b). This process of detrainment through
phase slips continues down to the smallest scale upon further weakening 
of the coupling.

The actual dynamics of phase slip initiation and propagation, particularly 
in the presence of runaway oscillators, is rather complicated and will be 
left for future investigation. It is, however, possible to gain some 
intuition about the factors governing 
the typical domain size through the following
consideration. In Fig.~\ref{fig:layer_3D}(b), for example, each of the 
light-colored regions can be considered as critical in the sense that a 
weaker $K$ would lead to fragmentation of the domain while a stronger $K$ 
would lead to merging with its neighbors. Let $\delta K$ be the 
increment in $K$ needed for the latter process to occur when the
typical size of light-colored domains is $\xi$. Upon synchronization of
two neighboring domains, their phase difference 
$\Delta\phi_0\sim \xi^{(4-d)/2}$ needs to be accommodated.
This is particularly so at the slip boundary, where bonds are turned
from being barely unstable to barely stable. The strength of these
bonds is of order $\delta K$, which offsets the phase gradient
$\Delta\phi_0/\xi\sim \xi^{(2-d)/2}$.
Let $K_c$ be the value of $K$ when $\xi=\infty$,
the above analysis yields a prediction for the size of synchronized domains
$\xi\sim \delta K^{-\nu}\simeq (K_{\rm c}-K)^{-\nu}$ at a given $K$,
where $\nu=\frac{2}{d-2}$.

\begin{figure}
\epsfxsize=\linewidth
\epsfbox{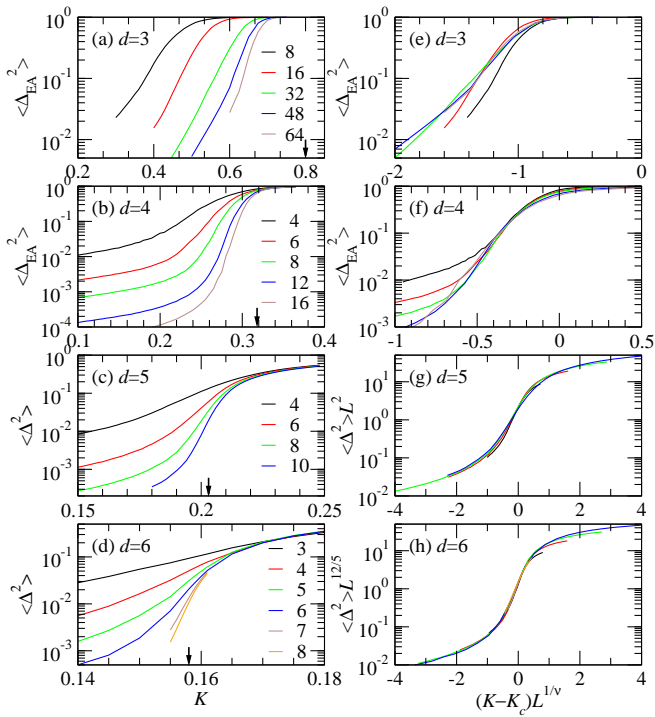}
\caption{(Color online) 
Order parameter against $K$ for various sizes
($L$ values given on each graph) in three to six dimensions.
Values of $K_{\rm c}$ [indicated by arrows in (a)-(d)] and $\nu$
used in the scaling plots are (e) $K_{\rm c}=0.8, \nu=2$; (f) $K_{\rm c}=0.318, \nu=1$;
(g) $K_{\rm c}=0.203, \nu=\frac{1}{2}$; (h) $K_{\rm c}=0.158, 
\nu=\frac{5}{12}$.}
\label{fig:finite_d_scaling}
\end{figure}

Figure~\ref{fig:finite_d_scaling} shows the entrainment order parameter
against $K$ for various system sizes. Thirty to several hundred samples 
were used to obtain the average in each case.
To verify the scaling predictions given above, we have analyzed the data
with the help of the usual finite-size scaling ansatz,
\begin{equation}
\langle\Delta^2\rangle = L^{-2\beta/\nu}\Phi(kL^{1/\nu}),
\label{eq:finite-size-scaling}
\end{equation}
where $k=(K-K_{\rm c})/K_{\rm c}$ and $\beta$ and $\nu$ are 
the order parameter and correlation length exponents, respectively.
For $d=5$ and 6, excellent data collapse is
achieved using $\beta=\frac{1}{2}$ and $\nu=\frac{\bar\nu}{d}=\frac{5}{2d}$ 
as in the globally-coupled case.
Interestingly, the scaling extends well into the detrained phase.
Noting that $\chi=N\langle\Delta^2\rangle$ corresponds to the susceptibility
in equilibrium magnetic systems, we conclude that the exponent
$\gamma=d\nu-2\beta=\frac{3}{2}$ for $d>4$ and differs from the globally
coupled case~\cite{ref:Daido90}.

For $d=3$ and 4, there is no convergence with increasing
system size (left panel) until $\langle\Delta_{\rm EA}^2\rangle$ reaches a 
value near one. This behavior is consistent with the idea that global 
entrainment takes place only when locally synchronized domains join to span 
the whole system. Overall, a good data collapse is achieved
in each case using the predicted exponents $\beta=0$ and $\nu=\frac{2}{d-2}$. 

In summary, through analytical arguments and large scale simulations
of the Kuramoto model in finite dimensions, 
we have established two types of critical behavior at
the onset of global entrainment. Above four dimensions, entrainment breaks the
global phase symmetry, as in the globally-coupled model, with identical scaling
exponents $\beta=\frac{1}{2}$ and $\bar\nu=\frac{5}{2}$. 
For $2<d\le 4$, and in particular for the physical dimension $d=3$,
global entrainment (detrainment) occurs via the aggregation
(fragmentation) of synchronized domains. The size of such domains
obeys scaling with an exponent $\nu=\frac{2}{d-2}$. 

This work was supported by the Korea Research Foundation Grant funded by the 
Korean Government (MOEHRD) (KRF-2006-331-C00123) (HH), by the Research Grants 
Council of the HKSAR (2017/03P and HKU3/05C) and by the Hong Kong 
Baptist University (FRG/01-02/II-65) (LHT).

\end{document}